\documentclass[journal]{IEEEtran}
\usepackage[T1]{fontenc}
\usepackage[utf8]{inputenc}
\usepackage[cmex10]{amsmath}
\usepackage{amssymb}
\usepackage{mathtools}
\usepackage{bbm}
\usepackage{graphicx}
\usepackage{cite}
\usepackage{enumitem}
\usepackage{tabularx}
\usepackage{booktabs}
\usepackage{makecell}
\usepackage{multirow}
\usepackage{array}
\usepackage{caption}
\usepackage{microtype}
\usepackage{url}

\DeclareCaptionFont{eightpt}{\fontsize{8pt}{9pt}\selectfont}
\begin{document}

\title{Multi-Agentic AI for Intent-Driven and Conflict-\\Aware rApp Policy Orchestration in Open RAN
}

\author{Haiyuan Li, Yulei Wu, Luyu Qi, Dimitra Simeonidou

\vspace{-0.4cm}

\thanks{This work was supported in part by the UK Engineering and Physical Sciences Research Council (EPSRC) grant EP/Y037243/1 and EP/X04047X/2 for the TITAN Telecoms Hub and grant EP/Y036514/1 for the JOINER project.}
 
\thanks{H. Li, Y. Wu, L. Qi, and D. Simeonidou are with the Smart Internet Lab, Department of Electrical and Electronic Engineering, University of Bristol, BS8 1QU, U.K. (e-mail: \{ocean.h.li, y.l.wu\}@bristol.ac.uk).}
}

\markboth{}{}

\maketitle 
\begin{abstract}
Open radio access network (O-RAN) enables a multi-vendor ecosystem that accelerates xApp development for near-real-time (RT) network management. However, this openness shifts a combinatorial coordination burden to the Non-RT layer, where rApp policies compose xApps from a growing catalog to realize operator intents. Coexisting intents further compound this challenge, as conflicts may arise among xApp actions, rApp policies, and their cross-layer interactions. To maximize intent admission while maintaining key performance indicator (KPI) compliance across incoming and incumbent intents, we propose a large language model-based Multi-Agentic AI framework. A Reasoning Agent screens candidate xApps against KPI contracts and produces a joint intent admission and xApp assignment decision. On post-deployment KPI violations, a Diagnosis Agent identifies the root causes of failure, labels the conflict mechanism, and guides the Reasoning Agent in subsequent attempts. Experimental results show that the framework based on GPT-OSS 120B achieves 28.6\% more KPI-compliant cases and admits over 15$\times$ as many incoming intents as state-of-the-art baselines. Under the Qwen 72B variant, Diagnosis increases compliant cases by 50.0\%, quadruples admitted intents, and reduces deployment attempts by 32.6\%. A live deployment conforming to O-RAN specifications further validates the feasibility of the proposed solution for conflict-aware rApp policy orchestration in real networks.
\end{abstract}

\begin{IEEEkeywords}
Open RAN, Agentic AI, rApp policy, intent-based networking, conflict mitigation
\end{IEEEkeywords}

\section{Introduction}
Mobile networks carry heterogeneous services with diverse performance and operational objectives. To address these demands, Open radio access network (O-RAN) exposes RAN control to third-party applications, expanding the ecosystem beyond vendor-specific functions~\cite{agarwal2025open, santos2025managing, li2025netmind}. In O-RAN, the non-real-time (Non-RT) RAN Intelligent Controller (RIC) within the Service Management and Orchestration (SMO) orchestrates rApps that translate operator intents into policies, while the near-real-time (Near-RT) RIC hosts xApps that carry out near-RT control such as resource scheduling, mobility optimization, and traffic steering~\cite{herrera2025tutorial, d2023orchestran, li2025toward, chen2025ran}. 
Although this layered open architecture enables flexible multi-vendor RAN control, it also introduces orchestration challenges. 
Compared with xApp development in the Near-RT RIC, automation at the Non-RT orchestration layer remains less mature. To serve diverse network management intents, the SMO must orchestrate rApps that select and compose xApps for each intent from a shared catalog without predefined xApp assignments. As the catalog grows, exhaustive profiling over the combinatorially expanding composition space becomes intractable. In addition, the coexistence of multiple intents further amplifies the risk of conflict. Coexisting intents may conflict through overlapping xApp actions, implicit runtime coupling across xApps, or incompatible rApp policies that cannot be jointly satisfied.
Therefore, automated Non-RT RAN management requires intent-driven and conflict-aware rApp policy orchestration over a shared xApp ecosystem.

Existing studies mainly address this challenge from two perspectives. 
The first is intent-based RAN management, which maps operator intents to executable policies, configurations, descriptors, or control loops through intent fulfillment and intent-driven control~\cite{banerjee2021intent,garces2024intent,mekrache2024llm,bimo2025intent,machado2026orion,nahum2025intent,habib2023intent,zhang2023intent,he2026agentic}.
The second focuses on conflict management, which detects and mitigates interference that arises when multiple rApp policies and their constituent xApps operate concurrently over shared RAN resources. Conflict management studies can be further organized by intervention time into pre-deployment and post-deployment methods. Pre-deployment methods screen or avoid conflicts ahead of execution through pre-execution screening, sandbox profiling, conflict-graph and causal learning, and optimization-based coordination~\cite{armstrong2024pre,delprever2025pacifista,al2026ran,sharma2025towards,mungari2025ran}. In comparison, post-deployment methods use runtime control messages, key performance indicator (KPI) observations, or network context to detect conflicts and coordinate xApp actions during operation~\cite{adamczyk2023conflict,wadud2024qacm,wadud2025xapp,varshney2025explainable,wadud2026ai,cinemre2025xapp}. 
Despite these advances, several limitations remain in Non-RT orchestration:

\vspace{-0.1cm}
\begin{itemize}[leftmargin=*]
    \item \textit{Lack of joint orchestration.} Intent-based RAN management reduces the burden of translating operator objectives into policies or configurations. However, most solutions consider single-intent settings and fixed intent categories without offering joint reasoning over intent admission, xApp assignment, and coordination between incumbent and incoming intents.

    \item \textit{Catalog-dependent profiling.}
    Existing orchestration and conflict-detection methods often rely on offline profiles, learned conflict graphs, sandbox measurements, or optimization models over predefined rApp policies. As the xApp catalog expands, exhaustive profiling of candidate compositions and manual maintenance of such policies become impractical.

    \item \textit{Overlooked cross-layer conflicts.}
    Current research predominantly targets Near-RT xApp interactions, especially overlapping RAN-parameter control. However, conflicts that originate at the rApp policy layer and manifest through xApp execution are rarely modeled.

    \item \textit{Disconnected pre- and post-deployment orchestration.}
    Pre-deployment methods focus on conflict avoidance but may miss issues that emerge only at runtime. Post-deployment methods focus on conflict mitigation but lack causal analysis to prevent recurrence. The two stages are not well integrated, leaving no closed-loop path from runtime KPI evidence back to orchestration decisions.
\end{itemize}

To address these limitations, we introduce a Multi-Agentic AI framework for intent-driven, conflict-aware rApp policy orchestration in O-RAN. The main contributions are summarized as follows:

\begin{itemize}[leftmargin=*]
    \item \textit{Multi-intent rApp policy orchestration formulation.}
    To jointly orchestrate coexisting intents over a combinatorially expanding xApp composition space while preventing cross-intent conflicts, we formulate rApp policy orchestration as a joint intent admission and xApp assignment problem with the objective of maximizing admitted incoming intents subject to post-deployment KPI compliance.
   
    \item \textit{Cross-layer conflict taxonomy.}
    We introduce a two-layer, five-class conflict taxonomy. The first layer inherits the Near-RT xApp conflict classes defined by O-RAN Alliance WG3~\cite{oran_conflict_mitigation}, covering direct attribute contention, indirect KPI coupling, and implicit runtime interference. The second layer extends to cross-layer conflicts between Non-RT rApp policies and Near-RT xApp execution, including shared-instance policy contention and inter-intent objective incompatibility. 

    \item \textit{Multi-agentic orchestration framework.}
    Building on the formulation, we propose a Multi-Agentic AI framework with guided chain-of-thought (CoT) designs for pre-deployment reasoning and post-deployment diagnosis. Following the taxonomy, the Reasoning Agent screens candidate xApps against KPI contracts and produces a joint intent admission and xApp assignment decision. On KPI violations after deployment, the Diagnosis Agent identifies the failed intent-KPI pairs, analyzes suspect xApps, classifies conflict mechanisms, and summarizes cross-attempt evidence to guide subsequent attempts, closing the loop between pre-deployment reasoning and post-deployment diagnosis.
    
    \item \textit{Benchmark and over-the-air testbed evaluation.}
    The framework is evaluated on a multi-intent O-RAN benchmark with 17 developed xApps and 19 operator intents across 10 incumbent/incoming-intent cases, with five runs per case for the full solution, ablations, and local large language model (LLM) variants. The proposed solution based on GPT-OSS 120B achieves 28.6\% more KPI-compliant cases and admits over 15$\times$ as many incoming intents as state-of-the-art baselines. The ablations and model comparison further show that guided CoT and LLM reasoning capability affect orchestration performance. Under the Qwen 72B variant, Diagnosis increases compliant cases by 50.0\%, quadruples admitted incoming intents, and reduces mean deployment attempts by 32.6\%. A live over-the-air proof of concept conforming to O-RAN specifications further validates the feasibility of the proposed Agentic AI solution for conflict-aware rApp policy orchestration in real networks.
\end{itemize}

The remainder of this paper is organized as follows.
Section~\ref{sec:literature} presents related work.
Section~\ref{sec:formulation} describes the orchestration scenario and formulates the optimization problem.
Section~\ref{sec:agentai} discusses the technical details of the multi-agent framework.
Section~\ref{sec:results} reports the benchmark evaluation, Section~\ref{sec:poc} presents the live-testbed proof of concept, and Section~\ref{sec:conclusion} concludes the paper.

\vspace{-0.3cm}
\section{Literature Review}
\label{sec:literature}

The expanding third-party xApp ecosystem introduces a heterogeneous set of candidate functions for realizing operator intents at the Non-RT layer. Moreover, conflicts among multiple intents may arise from incompatible KPI contracts, inconsistent rApp policies, or interfering xApp actions across the shared RAN infrastructure. 
Existing studies address these challenges mainly through intent-based RAN management and conflict management. The former translates high-level objectives into executable control, while the latter detects or mitigates conflicting interactions among deployed xApps and rApps. This section reviews both directions and identifies the need for intent-driven and conflict-aware rApp policy orchestration.

\vspace{-0.3cm}
\subsection{Intent-Based RAN Management}
\label{subsec:rw_intent}
Intent-based RAN management converts operator objectives into executable policies, configurations, descriptors, or control actions. Intent-fulfillment studies focus on the representation and translation stage. Banerjee et al. introduce an intent specification platform and an Intent Fulfillment System that translates formal intents into Cognitive Autonomous Network (CAN)-based RAN actions~\cite{banerjee2021intent}. Garcês et al. use a chatbot interface to translate natural-language commands into network policies enforced by an ONOS software-defined networking (SDN) controller~\cite{garces2024intent}. Mekrache and Ksentini use an LLM to generate network service descriptors for network function virtualization (NFV) deployment~\cite{mekrache2024llm}.

Beyond translation, intent-driven control optimizes the resulting network behavior through learned or agentic control loops. 
Radio scheduling, xApp orchestration, and slice resource allocation have been optimized with reinforcement learning under fixed intent objectives and action spaces~\cite{nahum2025intent,habib2023intent,zhang2023intent}. More recent studies extend intent-driven management with LLMs and agentic mechanisms, including distributed agents across rApp, xApp, and dApp timescales~\cite{elkael2026agentran}, RAN configuration generation from structured prompts and network feedback~\cite{bimo2025intent}, SMO-level semantic translation connected with Non-RT rApp and Near-RT xApp enforcement~\cite{machado2026orion}, and planning, tool-use, memory, and reflection for O-RAN management tasks~\cite{he2026agentic}. These studies reduce the manual cost of intent translation and control, but coexistence among intents, rApp policies, and independently developed xApps remains outside their main decision scope. A related conflict-oriented direction detects inconsistent control-parameter objectives or semantic conflicts among intents and policies~\cite{banerjee2022contradiction,de2024towards}. These conflict-oriented studies focus on semantic contradictions, while runtime network state, xApp catalog metadata, deployed rApp policies, and post-deployment KPI evidence remain largely absent.

\vspace{-0.3cm}
\subsection{Conflict Management in O-RAN}
\label{subsec:rw_conflict}
O-RAN conflict management is commonly grouped by intervention time. Pre-deployment methods detect or avoid conflicts before xApps act on the RAN. One group predicts configuration risk, as in Armstrong et al., who estimate KPI degradation from rApp-initiated configuration changes using performance-management, configuration, and KPI data~\cite{armstrong2024pre}. A second group profiles or learns conflict structure. Del Prever et al. introduce PACIFISTA, which profiles O-RAN applications in sandbox settings and builds application--attribute--key performance measurement (KPM) graphs to estimate conflict severity~\cite{delprever2025pacifista}. Zolghadr et al., Al Shami et al., and Sharma et al. use GraphSAGE, graph convolutional networks, and SHapley Additive exPlanations (SHAP) with causal inference to construct or explain xApp--attribute--KPI relationships~\cite{zolghadr2025learning,al2026ran,sharma2025towards}. Adjacent orchestration work optimizes how xApps are selected and shared for service realization. Mungari et al. propose OREO, which models RAN services as configurations of functions implemented by xApps and selects service configurations, xApp complexity levels, sharing decisions, and resource allocations under quality, latency, and resource constraints~\cite{mungari2025ran}. Zhang et al. and Erdol et al. reduce conflicts earlier during the development stage through team learning or xApp distillation~\cite{zhang2022team,erdol2025xapp}. These methods support risk screening, conflict-knowledge construction, resource-aware xApp placement, or conflict-aware xApp design, but their fidelity depends on offline profiles, learned graphs, predefined service configurations, resource estimates, or trainable xApps.

Post-deployment methods use runtime control messages, KPI observations, or network context after xApps have been deployed. At the control-message level, Adamczyk and Kliks propose a Near-RT RIC mitigation framework that intercepts xApp control messages to identify direct and indirect conflicts, and Sultana et al. demonstrate a control-message ordering mechanism on a software-defined radio O-RAN platform~\cite{adamczyk2023conflict,sultana2024software}. Parameter-level work then moves from arbitration to mitigation. Wadud et al. develop cooperative-bargaining and quality-of-service (QoS)-aware mechanisms for selecting feasible settings of contended parameters~\cite{wadud2024qacm,wadud2025xapp}. Beyond its pre-deployment use, explainable learning also appears in runtime diagnosis. Varshney et al. and Wadud et al. introduce models based on SHAP, graph neural networks (GNNs), and bidirectional long short-term memory (Bi-LSTM) for explaining or classifying runtime conflicts~\cite{varshney2025explainable,wadud2026ai}. Cinemre et al. and Giannopoulos et al. broaden runtime handling beyond single-parameter mitigation. Cinemre et al. propose a context-aware scheduler for active xApps without retraining them~\cite{cinemre2025xapp}, while Giannopoulos et al. introduce a workflow that combines proactive and reactive conflict handling with emulated-network validation of candidate xApp actions before live execution~\cite{giannopoulos2025comix}. These methods provide runtime observability, control-message arbitration, parameter-level mitigation, or xApp-level scheduling. However, the upstream Non-RT decision over intent admission and xApp assignment remains outside their decision scope.

\vspace{-0.2cm}
\section{System Model and Problem Formulation}
\label{sec:formulation}
\begin{figure}[t]
    \centering
    \includegraphics[width=1\linewidth]{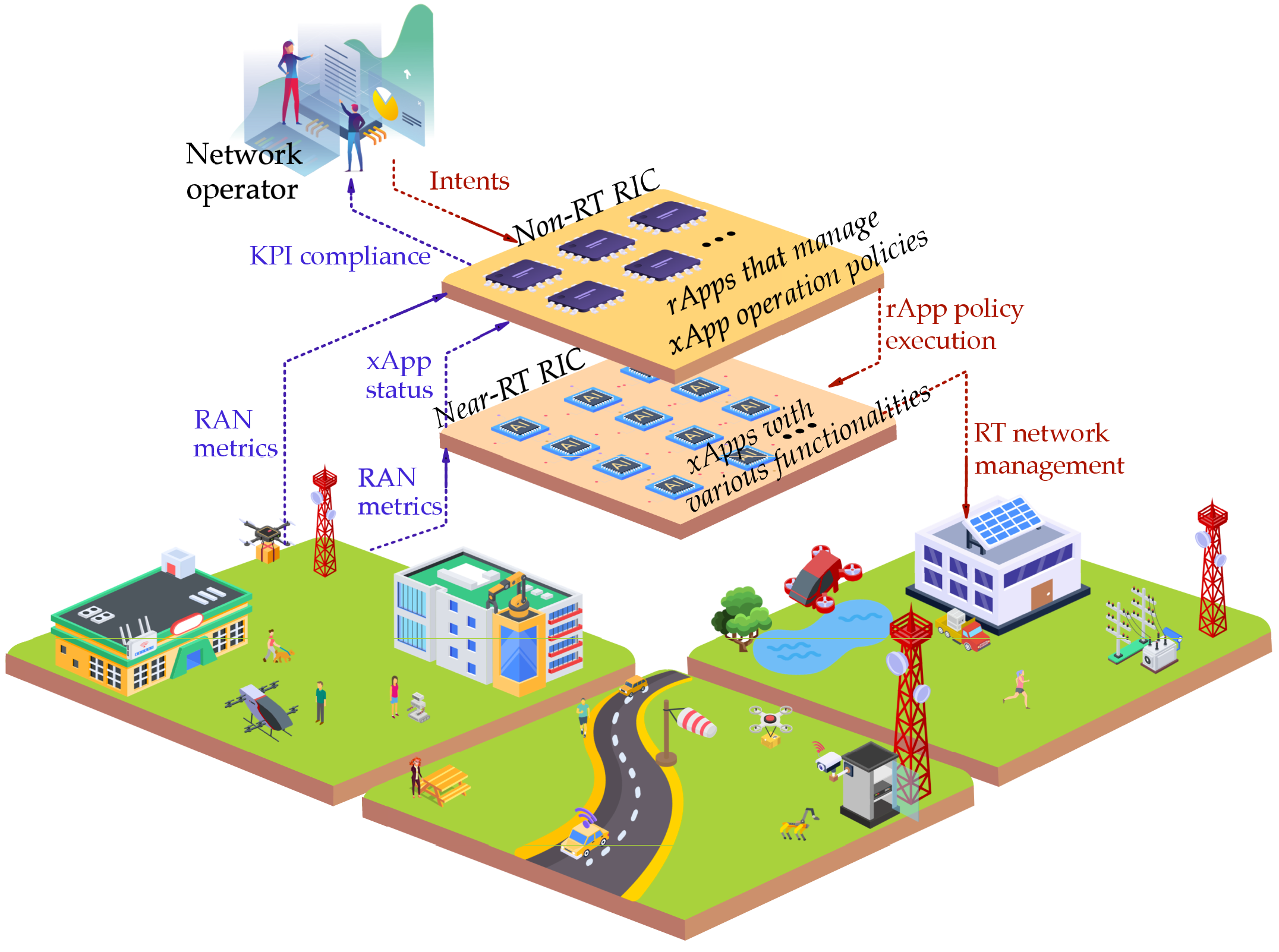}
    \caption{Multi-intent rApp policy orchestration across SMO, Non-RT RIC, Near-RT RIC, and RAN.}
    \label{fig:system}
    \vspace{-0.2cm}
\end{figure}

We consider a heterogeneous O-RAN deployment with three classes of entities. 
A set of access points (APs) $A$ spans multiple radio access technologies (RATs), including WiFi, 5G New Radio (NR), and LiFi.
A population of user equipment (UEs) $U$ generates service demands such as augmented reality/virtual reality (AR/VR) offloading, video streaming, and Internet of Things (IoT) telemetry. A set of multi-access edge computing (MEC) servers $E$ provides compute offloading and task scheduling.
UEs associate with APs over wireless links whose quality varies with mobility, interference, and load, while APs connect to MEC servers or the core network through backhaul links. Fig.~\ref{fig:system} illustrates the corresponding SMO, Non-RT RIC, Near-RT RIC, and RAN setting.

During network operation, operators may submit one or more incoming intents $I^{n}$ to the SMO, while a set of incumbent intents $I^{a}$ is already admitted and active in the network. 
The complete intent set under consideration is $I=I^{a}\cup I^{n}$. Each intent $i\in I$ includes an operator request or service-level objective targeting one or more KPIs, and each target KPI is associated with an explicit KPI contract, such as maintaining AR latency below 400\,ms or sustaining downlink throughput above 28\,Mb/s. An intent is represented as
\begin{equation}
    \begin{aligned}
    i = \bigl(
        &d_i,\,
        M_i,\,
        \{(\underline{g}_{i,m},\overline{g}_{i,m})\}_{m\in M_i}
        \bigr).
    \end{aligned}
    \label{eq:intent_tuple}
\end{equation}
The field $d_i$ is the operator request text or service-level description, and $M_i$ is the set of target KPIs of intent $i$. For each $m\in M_i$, the pair $(\underline{g}_{i,m},\overline{g}_{i,m})$ gives the acceptable KPI interval. One-sided KPI contracts are represented by setting the unused side of the interval to $-\infty$ or $+\infty$.

An admitted intent is realized by an rApp policy that is deployed in the Non-RT RIC and consists of one or more xApps selected from a shared xApp catalog $X$ maintained by the SMO. For each incumbent intent $i\in I^{a}$, the deployed rApp policy is denoted by $p_i\in\mathcal{R}$. Each xApp $x \in X$ is characterized by
\begin{equation}
x=\left(W_x,T_x,\Gamma_x\right),
\label{eq:xapp_tuple}
\end{equation}
where $W_x$ denotes the set of network attributes associated with xApp $x$ (e.g., AP selection weights, handover margins, and edge dispatch ratios), $T_x$ denotes the set of KPIs that can be affected by xApp $x$ (e.g., average latency, served throughput, and AP utilization), and $\Gamma_x$ denotes the runtime profile and validation metadata of the xApp, including its parameter schema and control scope.

\noindent\textbf{Definition 1 (rApp policy orchestration).}
Given incumbent intents $I^{a}$, incoming intents $I^{n}$, deployed rApp policies $\mathcal{R}$, and the xApp catalog $X$, rApp policy orchestration is the SMO decision that admits or rejects incoming intents and assigns xApps from $X$ to the rApp policy serving each admitted intent. The decision is represented by the admission variables $y_i$ and assignment variables $b_{i,x}$, with
\begin{equation}
    \begin{aligned}
    & y_i \in \{0,1\}, \quad i\in I,\\
    & b_{i,x} \in \{0,1\}, \quad i\in I,\ x\in X,\\
    & V \triangleq
    \{b_{i,x}\}_{i\in I,\,x\in X}
    \cup
    \{y_i\}_{i\in I}.
    \end{aligned}
    \label{eq:decision_space}
\end{equation}
Here $y_i=1$ admits intent $i$, $y_i=0$ rejects it, and $b_{i,x}=1$ assigns xApp $x$ to the rApp policy serving intent $i$. Each xApp assigned to one or more admitted intents has a single deployed instance. The active xApp set induced by $V$ is 
\begin{equation}
    X_{\mathrm{act}}(V)=\{x\in X \mid \sum_{i\in I} y_i b_{i,x}\ge 1\}.
\end{equation}
After deployment, the assigned xApps remain active in the Near-RT RIC and operate under their associated rApp policies, determining the RAN attribute trajectory and the resulting KPI observation $q_{i,m,t}(V)$ of intent $i$ on KPI $m$ at time $t$.

To evaluate compliance, each KPI contract defines an in-bound region
\begin{equation}
B_{i,m} = [\underline{g}_{i,m},\overline{g}_{i,m}], \quad m\in M_i.
\label{eq:kpi_bound_region}
\end{equation}
After a candidate rApp policy is deployed, KPI compliance is evaluated against $B_{i,m}$ over an evaluation window of duration $\theta_i$. Let $K_{i,m}=K_{i,m}(\theta_i)$ denote the set of observation instants for KPI $(i,m)$ in this window. The corresponding observations are aggregated into the window statistic
\begin{equation}
\widehat{q}_{i,m}(V) =
\frac{1}{|K_{i,m}|}
\sum_{t \in K_{i,m}}
q_{i,m,t}(V).
\label{eq:window_statistic}
\end{equation}
The KPI compliance verdict is
\begin{equation}
\eta_{i,m}(V) =
\mathbbm{1}\!\left[\widehat{q}_{i,m}(V) \in B_{i,m}\right].
\label{eq:kpi_compliance_verdict}
\end{equation}
Intent $i$ satisfies its KPI contract for $m$ when $\eta_{i,m}(V)=1$.

The objective is to optimize the admission and assignment variables in $V$ to admit as many incoming intents as possible while preserving the KPI compliance of incumbent intents and ensuring that every admitted incoming intent satisfies all of its KPI contracts.
Incumbent intents are assumed to be KPI-compliant under their deployed rApp policies $\mathcal{R}$ at the time of each orchestration decision. An incoming intent that cannot be jointly deployed is rejected in the current decision and may be reconsidered in a subsequent decision under an updated system state. The multi-intent orchestration problem is formulated as
\begin{subequations}\label{eq:problem}
\begin{gather}
\max_{V}
\quad
\sum_{i\in I^{n}} y_i
\label{eq:main_problem_obj}
\end{gather}
\vspace{-5mm}
\begin{align}
\hspace{2.8em}\text{s.t.} \quad
&
y_j = 1, \quad \forall j\in I^{a},
\label{eq:active_intent_keep} \\
&
b_{i,x} \le y_i, \quad \forall i\in I,\ \forall x \in X,
\label{eq:assignment_admission} \\
&
{\textstyle\sum_{x \in X}} b_{i,x} \ge y_i, \quad \forall i\in I,
\label{eq:assignment_support} \\
&
\eta_{i,m}(V) \ge y_i, \quad \forall i\in I,\ \forall m\in M_i.
\label{eq:compliance_constraint}
\end{align}
\end{subequations}
Together, Constraints~\eqref{eq:active_intent_keep} and~\eqref{eq:compliance_constraint} keep every incumbent intent admitted and preserve its KPI compliance under any new assignment. Constraints~\eqref{eq:assignment_admission} and~\eqref{eq:assignment_support} couple admission to assignment so that an admitted intent must be assigned at least one xApp and assigning an xApp to an intent requires its admission. The factor $y_i$ in~\eqref{eq:compliance_constraint} deactivates the compliance constraint for rejected incoming intents. The assignment of an incumbent intent is not fixed by~\eqref{eq:problem} and may be revised, provided the resulting assignment remains KPI-compliant.
Overall, the feasible action space grows exponentially, as each admitted intent can be served by any subset of the shared xApp catalog and multiple intents compete over $X$. Moreover, the realized $q_{i,m,t}(V)$ depends on semantic relationships among xApp descriptions, intent specifications, demand patterns, and inter-xApp interactions that cannot be captured by precomputed xApp profiles. This motivates the LLM-driven framework described in Section~\ref{sec:agentai}.

\vspace{-0.3cm}
\section{Agentic Solution for rApp Policy Orchestration}
\label{sec:agentai}
\begin{figure}[t]
    \centering
    \includegraphics[width=1\linewidth]{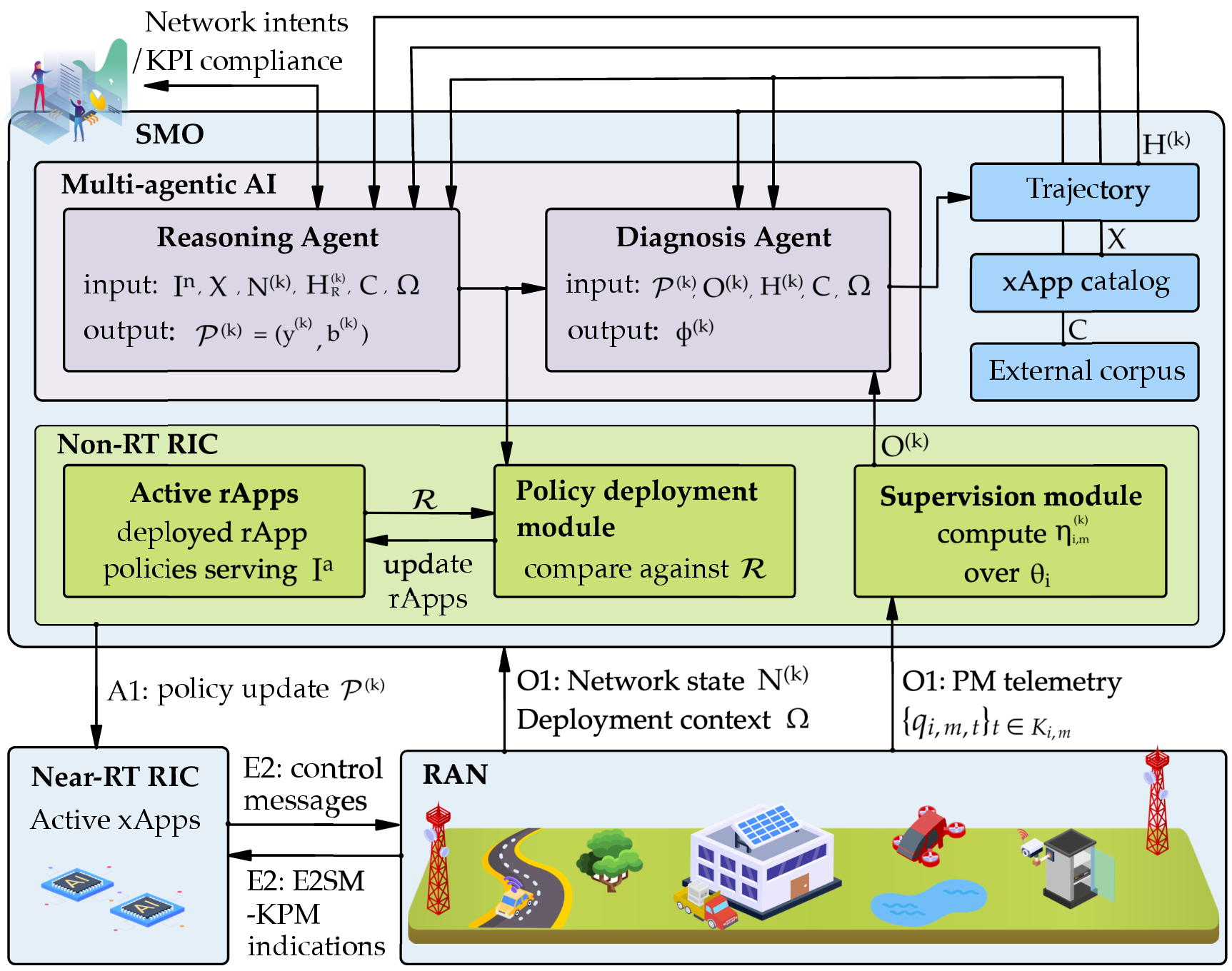}
    \caption{Closed-loop Agentic AI framework for rApp policy orchestration with reasoning, policy deployment, KPI supervision, diagnosis, trajectory update, and retrieved context.}
    \label{fig:framework}
\end{figure}

The orchestration problem~\eqref{eq:problem} couples incoming-intent admission and xApp assignment for rApp policy orchestration. In response, we propose a closed-loop Agentic AI framework as illustrated in Fig.~\ref{fig:framework}. It iterates pre-deployment reasoning, deployment, KPI supervision, and post-deployment diagnosis. Given incoming intents $I^{n}$ and incumbent intents $I^{a}$, the Reasoning Agent proposes a candidate rApp policy from the xApp catalog, incumbent deployment state, and retrieved O-RAN context. We introduce a policy deployment module that applies the candidate to the RAN and a supervision module that evaluates post-deployment KPI compliance against~\eqref{eq:compliance_constraint}. On KPI violations, the Diagnosis Agent converts the failure into structured evidence, and the framework launches another deployment attempt. We refer to one complete iteration of this loop as an \emph{orchestration cycle}. Attempts within a cycle are indexed by $k$ ($k=0$ for the initial attempt, $k\ge 1$ for retries). The retry budget $K_r$ limits the retries after the initial attempt, so a cycle contains at most $K_r+1$ deployment attempts. Both agents use guided CoT prompts that decompose each LLM call into ordered, labeled stages and record structured intermediate outputs before the final policy or attribution.

\vspace{-0.3cm}
\subsection{Conflict Taxonomy}
\label{sec:taxonomy}
To support pre-deployment screening and post-deployment diagnosis under a shared conflict vocabulary, we define five conflict mechanisms, denoted M1--M5 and grouped into two layers by their point of origin. Each mechanism describes an interaction structure that can cause an admitted intent to violate a KPI contract. The Reasoning Agent screens the mechanisms as hypotheses before deployment, and the Diagnosis Agent uses them as labels after supervision reports a violation. The first layer inherits the Near-RT xApp conflict classes defined by O-RAN Alliance WG3~\cite{oran_conflict_mitigation}, covering conflicts among xApps concurrently deployed in the Near-RT RIC. These include:
\begin{itemize}[leftmargin=*]
\item \emph{Direct conflict (M1).} Two distinct xApp instances act on a common network attribute ($W_x \cap W_{x'} \neq \varnothing$), and their contending writes on this attribute can cause an admitted intent to violate a KPI contract. The two instances belong to one intent or to separate intents.

\item \emph{Indirect conflict (M2).} Two xApps act on disjoint network attributes ($W_x \cap W_{x'} = \varnothing$), while their affected KPI sets overlap ($T_x \cap T_{x'} \neq \varnothing$) or are connected through a radio or edge-resource coupling, so their actions interfere at the KPI level.

\item \emph{Implicit conflict (M3).} Two or more xApps act on disjoint network attributes ($W_x \cap W_{x'} = \varnothing$) and declare disjoint affected KPI sets ($T_x \cap T_{x'} = \varnothing$), but their joint deployment still causes an admitted intent to violate a KPI contract. The violation arises from a runtime dependency outside the declared $W_x$ and $T_x$, such as a shared queue, AP load, or a control-loop timing effect.

\end{itemize}

The second layer covers cross-layer conflicts that originate from the interaction between rApp policies in the Non-RT RIC and the shared xApp catalog.

\begin{itemize}[leftmargin=*]
\item \emph{Policy conflict (M4).} A single xApp instance is shared by more than one admitted intent ($b_{i,x}=b_{j,x}=1$ for $i \neq j$). The xApp can support each intent under an isolated assignment, but cannot satisfy the attached rApp policies jointly under the current runtime network state. The contention in M4 originates from one shared xApp instance serving multiple rApp policies, whereas M1 involves distinct xApp instances contending over a common network attribute.

\item \emph{Objective incompatibility (M5).} Two admitted intents specify incompatible KPI contracts for the same target KPI ($M_i \cap M_j \neq \varnothing$) or for distinct but coupled KPIs that share radio or edge resources. The conflict remains under any xApp assignment, so the failure is rooted in the KPI contracts. Before deployment, M5 can be screened when KPI contracts are directly incompatible. For coupled KPIs, persistent failures across revised assignments provide post-deployment evidence of objective incompatibility.
\end{itemize}

The complete taxonomy is embedded in the guided prompts of both agents. The Reasoning Agent uses it to screen candidate assignments and guide policy revision, while the Diagnosis Agent uses it to classify post-deployment failures and organize cross-attempt evidence.

\vspace{-0.4cm}
\subsection{Reasoning Agent}
\label{sec:reasoning}
The Reasoning Agent produces a joint intent admission and xApp assignment decision to optimize the objective in~\eqref{eq:problem}. At attempt $k\ge 0$, the Reasoning Agent receives the incoming intent set $I^{n}$, the xApp catalog $X$, the runtime network observation $N^{(k)}$ (incumbent intents with their deployed rApp policies, active xApps, and online KPI measurements), the deployment context $\Omega$, the trajectory projection $H_R^{(k)}$, and the retrieved external context $C$. The deployment context $\Omega$ summarizes the access-technology topology, zones, and edge-server coverage and remains valid across attempts within one orchestration cycle. $H_R^{(k)}$ contains the prior policies and diagnosis attributions from the current orchestration cycle, while $C$ summarizes retrieved O-RAN knowledge. The full trajectory and retrieved context are defined in Section~\ref{sec:knowledge}. The agent uses one LLM call to parse the xApp profiles and the intent tuples~\eqref{eq:intent_tuple}, infer their KPI-level interactions, and produce the candidate rApp policy
\begin{equation}
\mathcal{P}^{(k)} = (y^{(k)},\, b^{(k)}) = f_1\!\left(I^{n},\,X,\,N^{(k)},\,\Omega,\,H_R^{(k)},\,C\right),
\label{eq:reasoning_call}
\end{equation}
where $f_1$ denotes the LLM call over the inputs above. The pair $(y^{(k)},b^{(k)})$ determines the variable realization $V^{(k)}$ used for compliance evaluation. Under the guided CoT design, the Reasoning Agent decomposes its LLM call into two ordered stages, R1 and R2. Their intermediate outputs are recorded as labeled structured fields alongside the final policy.

\emph{R1 (perception and structural screening)} parses the inputs and constructs candidate xApp subsets $\mathcal{X}_i\subseteq X$ that satisfy $\bigcup_{x\in \mathcal{X}_i}T_x\supseteq M_i$ and whose validated KPI effects are consistent with the corresponding contracts for each incoming intent $i\in I^{n}$. Each candidate is screened against the incumbent intent state and the candidates of the other incoming intents using the mechanisms that can be checked before deployment. The screen checks M1 through network-attribute overlap between candidate xApps and active xApps, and M2 through affected-KPI overlap or known radio and edge couplings under disjoint network attributes. Under M5, R1 excludes an intent only when its KPI contracts or a calibrated KPI feasibility audit already rule out a compliant joint assignment. The screen treats M3 and M4 as risks and keeps the associated candidates eligible. M3 depends on runtime coupling that structural signals cannot expose, and M4 remains a shared-instance risk until supervision shows that the shared instance cannot jointly satisfy the attached rApp policies. For attempts $k\ge 1$, R1 also uses $H_R^{(k)}$ to exclude previously failed assignments, xApps marked as persistent suspects, and ineffective assignment changes reported by the Diagnosis Agent. It returns the screened candidate sets and identifies any incumbent intents whose current assignments may require revision to preserve KPI compliance.

\emph{R2 (joint xApp assignment)} selects the joint decision $(y^{(k)},\,b^{(k)})$ over the screened candidate sets from R1. The selection targets the admission objective in~\eqref{eq:main_problem_obj} and keeps every incumbent intent admitted and KPI-compliant. Incumbent assignments belong to the assignment space. Decisions admitting the same number of incoming intents are equivalent under the objective in~\eqref{eq:main_problem_obj}. R2 therefore breaks ties by retaining the current assignment of each incumbent intent by default, revising it only when R1 indicates that reassignment is needed, and requiring any revised assignment to satisfy the compliance constraint of that intent. An incoming intent $i$ is rejected when R1 produces an empty candidate set for $i$ or when the most recent Diagnosis attribution reports a persistent failure for $i$. A rejected intent receives $y_i^{(k)}=0$ and $b_{i,x}^{(k)}=0$ for all $x\in X$. R2 returns the admission decisions, the xApp assignments, and a per-intent rationale. For attempts $k\ge 1$, the rationale references the prior Diagnosis attribution, and R2 uses its mechanism labels and trajectory evidence to revise the assignment under the conflict taxonomy in Section~\ref{sec:taxonomy}.

\vspace{-0.3cm}
\subsection{Diagnosis Agent}
\label{sec:diagnosis}
The retry mechanism relies on structured failure evidence $\phi^{(k)}$ for subsequent attempts. The \emph{Diagnosis Agent} produces this evidence and is invoked only after a deployed candidate fails KPI supervision. For the failed attempt $k$, it maps the deployed policy $\mathcal{P}^{(k)}$, the observation record $O^{(k)}$ emitted by the supervision module, the within-cycle trajectory $H^{(k)}$, the retrieved context $C$, and the deployment context $\Omega$ to the attribution
\begin{equation}
\phi^{(k)} = f_2\!\left(\mathcal{P}^{(k)},\,O^{(k)},\,H^{(k)},\,C,\,\Omega\right),
\label{eq:diagnosis_call}
\end{equation}
where $f_2$ denotes the LLM call made by the Diagnosis Agent. Under the guided CoT design, the attribution is organized into four ordered stages, D1--D4, that move from the affected KPI surface to suspect xApps, mechanism labels, and cross-attempt patterns.

\emph{D1 (affected surface)} identifies the intent-KPI pairs whose compliance condition was violated. The affected set is
\begin{equation}
F^{(k)} = \bigl\{(i,m) \mid y_i^{(k)}=1,\; \eta_{i,m}(V^{(k)}) = 0\bigr\}.
\label{eq:diagnosis_layer1}
\end{equation}
This set is read directly from the supervision output and is nonempty by construction.

\emph{D2 (suspect xApp set)} narrows each $(i,m)\in F^{(k)}$ to a suspect set $Z_{i,m}^{(k)}\subseteq X$ of xApps implicated in the violation. An xApp enters the suspect set through three signals from the current attempt. A structural signal marks xApps whose writable attributes $W_x$ are coupled to the affected KPI. A runtime signal marks xApps whose write events align temporally with the observed KPI degradation. A catalog signal marks xApps whose declared affected KPI set $T_x$ contains the violated metric. D2 leaves the suspect set empty when these signals fail to discriminate among candidates.

\emph{D3 (conflict mechanism)} identifies the conflict mechanism among the suspects, drawn from the taxonomy in Section~\ref{sec:taxonomy}, and records it as a mechanism label $L_{i,m}^{(k)}$ for each supported affected pair. The label gives the next Reasoning attempt a mechanism-level explanation of the observed failure. Post-deployment telemetry further allows D3 to confirm M3 and M4, which the structural screen in R1 treats only as risks. D3 leaves the label unassigned when the evidence supports no single mechanism.

\emph{D4 (cross-attempt synthesis)} reads the trajectory $H^{(k)}$, consisting of the prior policies, telemetry records, and attributions, and reports persistent patterns that are not visible from the current attempt alone. The agent issues a set of structured observations $\chi^{(k)}$, each describing one cross-attempt regularity together with the supporting trajectory evidence. The implementation uses three recurring categories. (i) \emph{Persistent failure} denotes an intent-KPI pair $(i,m)$ that appears in the affected set $F^{(j)}$ across attempts with different assignment decisions. The recurrence supplies evidence against admitting intent $i$ again under the current conditions. (ii) \emph{Persistent suspect} denotes an xApp $x$ that appears in the suspect sets $Z^{(j)}_{i,m}$ across multiple attempts and across different intent-KPI pairs. D4 marks this xApp as ruled out for subsequent attempts. This mark extends the single-attempt suspicion recorded by D2. (iii) \emph{Ineffective perturbation} denotes a class of assignment changes that failed to shrink the affected set across attempts. D4 records the change as ineffective, allowing the next Reasoning attempt to avoid repeating it and instead select a different xApp assignment or reject the incoming intent.
D1, D2, and D3 analyze the most recent failed attempt, whereas D4 compares evidence across attempts. D4 is applied from the second attempt onward and reports a pattern only when evidence from at least two attempts supports it. No additional repetition threshold is imposed beyond this minimum evidence requirement.

\vspace{-0.4cm}
\subsection{Trajectory and External Knowledge}
\label{sec:knowledge}
Both agents consume two context sources beyond the current intent and xApp inputs. The within-cycle trajectory $H$ records the attempts of the current orchestration cycle, and the agents read it to ground retry decisions in prior failure evidence. The retrieved context $C$ supplies O-RAN terminology, architectural constraints, and deployment knowledge for reasoning in the O-RAN setting.

\subsubsection{Within-Cycle Trajectory}
The trajectory accumulates per-attempt records within one orchestration cycle. After a failed attempt $k$, the supervision module emits the telemetry record $O^{(k)}$, the Diagnosis Agent produces the attribution $\phi^{(k)}$, and the retry loop appends both to the trajectory as
\begin{equation}
H^{(k+1)} = H^{(k)} \mathbin{\Vert} \bigl((\mathcal{P}^{(k)},\,O^{(k)},\,\phi^{(k)})\bigr),\quad H^{(0)} = \emptyset.
\label{eq:trajectory_update}
\end{equation}
The resulting trajectory is supplied to the next attempt as an ordered structured field. Its size grows linearly with the attempt count and is bounded by $K_r+1$ deployment attempts.

From this accumulated record, the two agents read different projections. The Reasoning Agent receives the ordered projection
\begin{equation}
H_R^{(k)} \triangleq \bigl((\mathcal{P}^{(j)},\,\phi^{(j)})\bigr)_{j<k},
\label{eq:reasoning_trajectory_projection}
\end{equation}
consisting of the prior policies and their attributions. This projection omits the telemetry records because the attribution summarizes the failure evidence needed for re-planning. The Diagnosis Agent receives the full trajectory with the telemetry records, since cross-attempt comparison of write traces and KPI patterns underpins the pattern synthesis in D4. The trajectory is retained only for the duration of its cycle. A KPI-compliant attempt terminates the cycle without another trajectory update. If the cycle instead reaches $K_r+1$ attempts, incumbent intents continue under their prior rApp policies $\mathcal{R}$, and every unserved incoming intent is recorded as $y_i=0$. 

\subsubsection{External Knowledge via Retrieval}
The framework augments agent reasoning with retrieval over an external corpus of O-RAN technical documentation, including working-group specifications for xApp interfaces and procedural guidance on coexistence handling. The corpus is segmented into passages $\{d_p\}$, and each passage is embedded as $\zeta_p = \psi(d_p)$ by the embedding model $\psi(\cdot)$. Each agent issues a task-specific retrieval query. The Reasoning Agent retrieves xApp interface descriptions and documented coexistence patterns using the parsed intent fields and current network state. The Diagnosis Agent retrieves conflict-mechanism descriptions using the affected pairs $F^{(k)}$ and suspect sets $Z_{i,m}^{(k)}$. Candidate passages are ranked separately by BM25 lexical matching and embedding similarity, and the two rankings are combined using reciprocal rank fusion (RRF). Maximal marginal relevance then selects a non-redundant top-$L_{\mathrm{rag}}$ subset as the retrieved context $C$, which is inserted into the prompt of the requesting agent~\cite{lewis2020retrieval}. The corpus persists across orchestration cycles and is the only state shared among them.

\vspace{-0.2cm}
\subsection{Non-LLM Modules}
\label{sec:modules}
Beyond the two agents, the closed-loop framework includes two deterministic modules that handle policy deployment and KPI supervision in the Non-RT RIC.
The \emph{policy deployment module} receives the candidate policy $\mathcal{P}^{(k)}=(y^{(k)},b^{(k)})$ and derives the active xApp set $X_{\mathrm{act}}^{(k)}=X_{\mathrm{act}}(y^{(k)},b^{(k)})$. It then compares the candidate policy with the deployed rApp policies $\mathcal{R}$ to determine newly required xApps, updated xApp assignments, and unchanged rApp policies. Newly required xApps are instantiated through the O1 and O2 interfaces, and xApp assignments are dispatched through the A1 interface. For an incumbent intent, the module updates only the assignment entries $b_{i,x}$ of its deployed rApp policy, since the admission variable is fixed by~\eqref{eq:active_intent_keep}. Retained xApps and unchanged rApp policies continue without redeployment.
The \emph{supervision module} evaluates KPI compliance after each deployment. It collects O1 performance-management telemetry and E2SM-KPM indications, aggregates the measurements over the evaluation window $\theta_i$, and computes the KPI compliance verdict $\eta_{i,m}$ in~\eqref{eq:kpi_compliance_verdict}. The resulting observation record is emitted as $O^{(k)}$. If $\eta_{i,m}=0$ for any target KPI $m\in M_i$ of an admitted intent, the module marks the deployment as non-compliant and invokes the Diagnosis Agent. Otherwise, the deployment is deemed KPI-compliant and terminates the current orchestration cycle.

The framework above establishes a closed-loop methodology for rApp policy orchestration. The optimization problem in~\eqref{eq:problem} specifies joint incoming-intent admission and xApp assignment under fixed KPI contracts. In operational deployments, the network operator may iteratively refine KPI contracts and specify admission priorities through interaction with the agentic interface, such as by designating mandatory intents subject to feasibility and incumbent-intent compliance. These operator inputs can be incorporated into the optimization formulation and guided prompts without retraining the LLM, while interactive requirement negotiation and priority-aware orchestration remain outside the present scope.

\vspace{-0.2cm}
\section{Experimental Evaluation}
\label{sec:results}

To evaluate the effectiveness of the proposed closed-loop Multi-Agentic AI framework for rApp policy orchestration, we measure incoming-intent admission while requiring post-deployment KPI compliance for incumbent and admitted incoming intents. In a heterogeneous O-RAN environment, all methods use the same RAN realization, xApp catalog, evaluated intent set, and supervision module. For comparison, we adapt three baselines from the literature: OREO~\cite{mungari2025ran}, PACIFISTA~\cite{delprever2025pacifista}, and Causal-SHAP~\cite{sharma2025towards}. We also evaluate two internal ablations and an alternative local LLM backend.

\begin{table}[!t]
\centering
\caption{RAN setup and network attributes.}
\label{tab:experiment_control_space}
\scriptsize
\setlength{\tabcolsep}{3pt}
\renewcommand{\arraystretch}{1.12}
\begin{tabularx}{\columnwidth}{@{}
>{\raggedright\arraybackslash}p{2.15cm}
>{\raggedright\arraybackslash}X
@{}}
\toprule
\textbf{Item} & \textbf{Setting} \\
\midrule
Deployment &
Heterogeneous RAN with three 5G, five WiFi, and four LiFi APs, 300 mobile UEs, AP-to-edge wired backhaul, and four MEC servers. \\

Spatial and temporal scale &
$600\times400$\,m planar deployment area. Control step $\Delta t=0.2$\,s. \\

Wireless parameters &
5G/WiFi/LiFi bandwidths $80/60/160$\,MHz. Transmit powers $43/23/28$\,dBm. Maximum serving rates $R_a^{\max}$ of $1200/600/600$\,Mb/s. Wall loss $14$\,dB. Noise floor $-85$\,dBm. \\

Channel and smoothing &
Path-loss exponent $2.4$. Shadowing standard deviation $5$\,dB. 5G spectral-efficiency ceiling $7.4$\,b/s/Hz. Exponentially weighted moving average (EWMA) factor $0.7$ for stability-sensitive KPIs. \\

Edge and backhaul parameters &
MEC servers with $400/600/900/1250$\,GFLOPS compute budgets and $600/1000/1800/2400$\,Mb/s wired links. \\

KPI evaluation window &
Compliance evaluated over $\theta_i=120$\,s. \\

\midrule
Radio/access attributes &
$c_{\mathrm{5G}}^{\mathrm{cap}}\in[0.3,3.0]$ scales 5G serving capacity.
$b_{\mathrm{5G}}\in[-6,6]$\,dB biases 5G cell selection.
$w_{\mathrm{5G}}\in[0.5,2.0]$ is the 5G scheduling priority weight.
$s_{\mathrm{5G}},s_{\mathrm{W}}\in[0,0.6]$ are 5G/WiFi sleep fractions.
$H^{\mathrm{ho}}\in[0,6]$\,dB is the handover hysteresis margin. \\

Access-policy attributes &
$\pi^{\mathrm{RAT}}\in\{\text{balanced},\text{5G},\text{WiFi},\text{LiFi}\}$ selects the RAT preference.
$\varphi^{\mathrm{sel}}\in\{\text{max-rate},\text{score},\text{nearest}\}$ selects the AP/cell association rule. \\

Edge/compute attributes &
$\alpha^{\mathrm{adm}}\in[0.4,1.0]$ is the per-server edge offload admission probability.
$\nu^{\mathrm{cap}}\in[0.3,3.0]$ scales edge compute capacity.
$\lambda^{\mathrm{cons}}\in[0,1.0]$ controls workload consolidation strength.
$\nu^{\mathrm{cpu}}\in[0.3,2.5]$ scales edge CPU frequency. \\

UE/offload attributes &
$\delta^{\mathrm{UE}}\in[0.3,1.0]$ is the UE active-time fraction.
$\omega^{\mathrm{off}}\in[0.3,1.0]$ biases UE offloading aggressiveness. \\

Reliability attribute &
$p^{\mathrm{dup}}\in[0,1.0]$ is the packet-duplication probability for reliability-critical traffic. \\
\bottomrule
\end{tabularx}
\end{table}

\begin{table*}[!t]
\centering
\caption{xApp catalog used for rApp policy orchestration.}
\label{tab:xappcatalog}
\scriptsize
\setlength{\tabcolsep}{2.2pt}
\renewcommand{\arraystretch}{1.12}
\begin{tabularx}{\textwidth}{@{}
l
l
c
>{\raggedright\arraybackslash}p{0.25\textwidth}
>{\raggedright\arraybackslash}X
@{}}
\toprule
\textbf{xApp} &
\textbf{Alg.} &
\textbf{Attributes} &
\textbf{Affected KPI Direction} &
\textbf{xApp Action} \\
\midrule
X-01 & RF &
$\delta^{\mathrm{UE}}$, $\omega^{\mathrm{off}}$ &
Reduce UE energy consumption &
Reduces UE active time and offloading when predicted energy savings outweigh service loss.\\

X-02 & RF &
$s_{\mathrm{5G}}$, $s_{\mathrm{W}}$ &
Reduce aggregate access-point energy &
Schedules sleep mode for underutilized 5G/WiFi APs while preserving active-user capacity. \\

X-03 & RF &
$\nu^{\mathrm{cpu}}$ &
Reduce per-server edge energy &
Lowers edge-server CPU frequency when latency and demand allow lower-power execution. \\

X-04 & RF &
$\lambda^{\mathrm{cons}}$ &
Reduce total edge-infrastructure energy &
Consolidates workloads across edge servers so lightly loaded servers enter low-power states. \\

X-05 & RF &
$c_{\mathrm{5G}}^{\mathrm{cap}}$ &
Increase delivered downlink throughput &
Expands effective 5G radio headroom when demand pressure limits user throughput. \\

X-06 & RF &
$\pi^{\mathrm{RAT}}$, $\varphi^{\mathrm{sel}}$ &
Increase delivered downlink throughput &
Selects RAT and AP association policy by predicting the impact of WiFi/5G/LiFi rules. \\

X-07 & LinUCB &
$b_{\mathrm{5G}}$ &
Improve throughput-aware load balance &
Adjusts 5G cell-selection bias to reduce congestion under load imbalance. \\

X-08 & MLP &
$w_{\mathrm{5G}}$ &
Increase scheduled throughput &
Sets 5G scheduler priority so throughput-oriented traffic receives more resources under contention. \\

X-09 & DQN &
$\nu^{\mathrm{cpu}}$ &
Reduce AR end-to-end latency &
Raises edge compute speed for latency-sensitive flows to drain queues and reduce latency. \\

X-10 & RF &
$\nu^{\mathrm{cap}}$ &
Reduce AR end-to-end latency &
Adds effective edge compute capacity when processing headroom is the dominant bottleneck. \\

X-11 & DQN &
$\omega^{\mathrm{off}}$, $\pi^{\mathrm{RAT}}$ &
Reduce AR end-to-end latency &
Adjusts UE offloading preference and RAT selection to favor lower-latency execution paths. \\

X-12 & RF &
$w_{\mathrm{5G}}$ &
Reduce AR end-to-end latency &
Shifts radio scheduler priority toward latency-sensitive traffic under radio contention. \\

X-13 & QGBDT &
$\alpha^{\mathrm{adm}}$ &
Improve conservative edge-admission utility &
Sets offload admission aggressiveness using lower-quantile utility estimates to avoid overload. \\

X-14 & MLP-G &
$w_{\mathrm{5G}}$ &
Increase task-completion success &
Sets a guarded 5G scheduling weight for task services while avoiding unsafe extreme actions. \\

X-15 & MLP &
$\nu^{\mathrm{cap}}$ &
Increase task-completion success &
Raises effective edge-service capacity for task traffic under compute bottlenecks. \\

X-16 & RF &
$p^{\mathrm{dup}}$ &
Increase transmission success probability &
Enables selective packet duplication when reliability gains justify the extra radio load. \\

X-17 & MLP-G &
$H^{\mathrm{ho}}$, $b_{\mathrm{5G}}$ &
Reduce handover ping-pong &
Adjusts mobility hysteresis and cell-selection bias to suppress ping-pong without excessive stickiness. \\
\bottomrule
\end{tabularx}
\end{table*}

\begin{figure*}[!t]
\centering
\includegraphics[width=\textwidth]{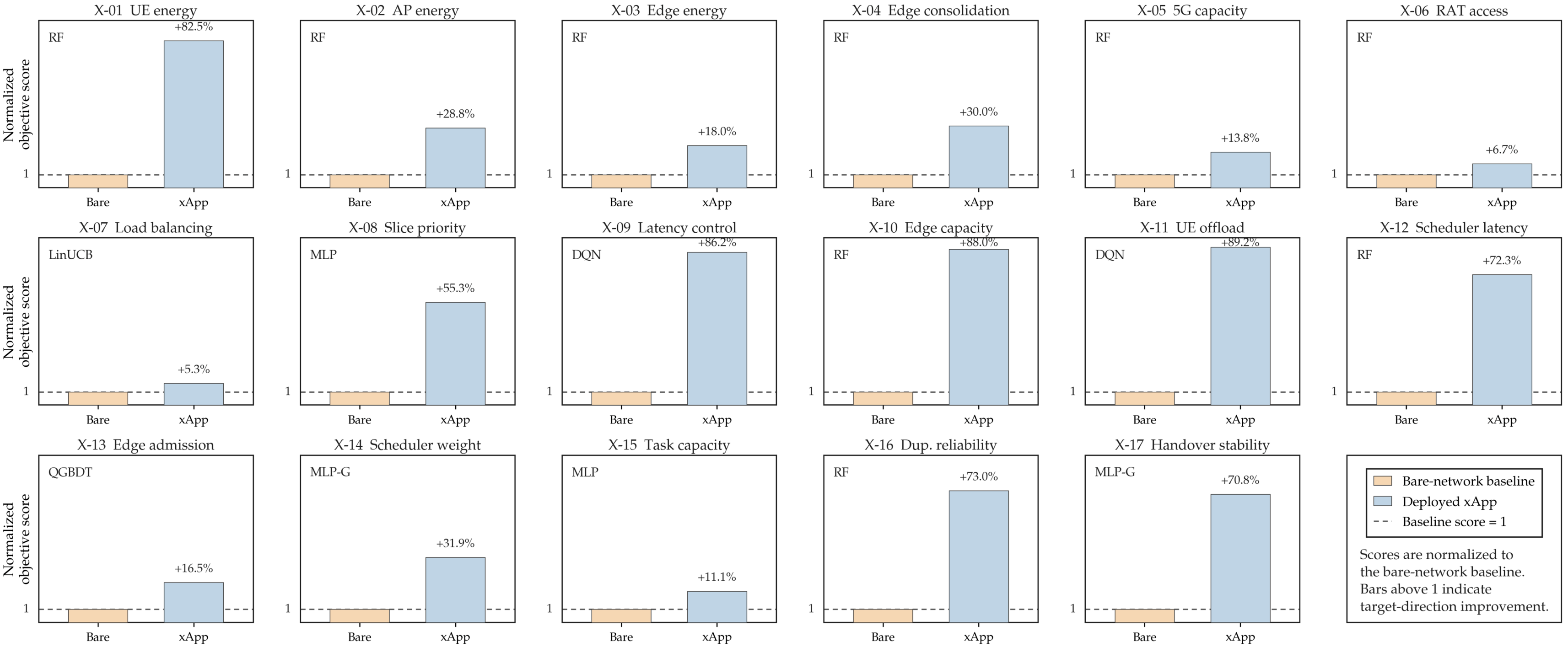}
\caption{xApp-level validation of the 17 xApps against their target KPI directions.}
\label{fig:xapp_controller_validation}
\vspace{-0.1cm}
\end{figure*}

\vspace{-0.4cm}
\subsection{RAN and Control Space}
\label{sec:results:environment}

Following the system model in Section~\ref{sec:formulation}, the evaluation uses a heterogeneous O-RAN realization with multi-RAT access, edge compute, and energy accounting. The access network contains 5G, WiFi, and LiFi APs serving mobile UEs. The edge layer contains finite MEC queues and wired backhaul links, with admission and compute-scaling network attributes controlled by xApps. The supervision module monitors downlink throughput, AR latency, task success, transmission success, ping-pong rate, and the AP, edge-server, and UE energy metrics tracked by the energy model.

The edge-compute layer maps offloaded demand to MEC queues. Each edge server $e\in E$ has configured compute capacity $C_e$ and effective capacity $\widehat{C}_{e}^{(t)}=C_e\nu_{e}^{\mathrm{cpu},(t)}\nu_{e}^{\mathrm{cap},(t)}$, where $\nu^{\mathrm{cpu}}_e$ and $\nu^{\mathrm{cap}}_e$ are the CPU-frequency and capacity-scaling controls. If $\Lambda_e^{(t)}$ is the admitted GFLOP demand arriving during step $t$, the queue state $Q_e^{(t)}$ evolves by standard first-in-first-out accumulation under a service rate $\widehat{C}_e^{(t)}$. For an offloaded task $\tau$, the end-to-end latency is computed as
\begin{equation}
T_{\tau}^{(t)}=T_{\tau}^{\mathrm{air}}+T_{\tau}^{\mathrm{bh}}+\frac{Q_e^{(t)}}{\widehat{C}_e^{(t)}}+\frac{c_\tau}{\widehat{C}_e^{(t)}}+T_e^{\mathrm{wake}},
\label{eq:env:latency}
\end{equation}
where $T_{\tau}^{\mathrm{air}}$ and $T_{\tau}^{\mathrm{bh}}$ are the air-interface and backhaul delays, the third and fourth terms are the queueing and processing delays at the serving edge server, $c_\tau$ is the task compute demand, and $T_e^{\mathrm{wake}}$ is the wake-up delay associated with the edge sleep state. The per-server admission control $\alpha^{\mathrm{adm}}_e$ rejects a fraction $1-\alpha^{\mathrm{adm}}_e$ of offload requests before they enter the queue of server $e$.

AP and edge energy follow static-plus-utilization-proportional models with sleep-state multipliers. AP sleep controls reduce active radio availability and dynamic AP power, and edge sleep sets admitted edge utilization to zero. The capacity-scaling control $\nu_e^{\mathrm{cap}}$ changes schedulable edge capacity, while $\nu_e^{\mathrm{cpu}}$ is coupled directly to dynamic edge power. UE energy is computed from idle time, radio association time, and per-RAT transmission activity.
The RAN parameters are summarized in Table~\ref{tab:experiment_control_space}.

xApps act on the access network through the association, scheduling, sleep, and capacity attributes in
\begin{equation}
\begin{aligned}
\mathbf{u}^{(t)} = (&c_{\mathrm{5G}}^{\mathrm{cap}}, b_{\mathrm{5G}}, w_{\mathrm{5G}}, s_{\mathrm{5G}}, H^{\mathrm{ho}},
\pi^{\mathrm{RAT}}, \varphi^{\mathrm{sel}}, s_{\mathrm{W}},\\
&\alpha^{\mathrm{adm}}, \nu^{\mathrm{cap}}, \lambda^{\mathrm{cons}}, \nu^{\mathrm{cpu}},
\delta^{\mathrm{UE}}, \omega^{\mathrm{off}}, p^{\mathrm{dup}}).
\end{aligned}
\label{eq:env:control_vector}
\end{equation}
On the radio side, access network attributes controlled by xApps affect the throughput KPI through capacity scaling, association, and scheduling. The radio model first computes a per-link rate ceiling from transmit power, environment-determined channel gain, noise, interference, bandwidth, and a clipped spectral-efficiency map. The aggregate AP serving capacity is capped by $R_a^{\max}$ and scaled by the capacity and sleep attributes. Association controls determine the serving AP of each UE, and the scheduling weight $w_{\mathrm{5G}}$ allocates the resulting AP capacity among associated UEs. The downlink-throughput KPI is computed after this allocation.

\begin{table*}[t]
\centering
\caption{Evaluated operator intents and KPI contracts used in the benchmark.}
\label{tab:intentcatalog}
\scriptsize
\renewcommand{\arraystretch}{1.14}

\begin{tabularx}{\textwidth}{@{}lXl@{}}
\toprule
\textbf{Intent} &
\textbf{Operator request} &
\textbf{KPI contract} \\
\midrule
I-01 & Keep high-demand video users above the downlink throughput contract during heavy radio load. & Downlink throughput $\geq$ 28\,Mb/s \\
I-02 & Maintain stable media-stream throughput when radio congestion increases. & Downlink throughput $\geq$ 28\,Mb/s \\
I-03 & Preserve AR navigation throughput for indoor mobile users across WiFi, LiFi, and 5G access. & Downlink throughput $\geq$ 28\,Mb/s \\
I-04 & Keep interactive AR end-to-end latency within the service bound under edge offloading. & AR latency $\leq$ 400\,ms \\
I-05 & Guarantee tighter AR latency for real-time control traffic with MEC processing. & AR latency $\leq$ 300\,ms \\
I-06 & Keep remote-control tasks completing successfully when MEC queues are loaded. & Task success ratio $\geq$ 0.27 \\
I-07 & Maintain deadline completion for industrial control tasks under shared edge resources. & Task success ratio $\geq$ 0.27 \\
I-08 & Maintain packet-delivery reliability for critical traffic under radio contention. & Transmission success $\geq$ 0.6 \\
I-09 & Suppress ping-pong handovers for mobility-sensitive UEs moving across cells. & Ping-pong rate $\leq$ 0.05\,s$^{-1}$ \\
I-10 & Reduce aggregate AP energy while maintaining service during low-load periods. & AP energy $\leq$ 260\,J/window \\
I-11 & Keep each MEC server within its edge-energy budget under offloaded demand. & Edge energy $\leq$ 58\,J/server/window \\
I-12 & Reduce aggregate MEC-server energy by consolidating edge workloads across available servers. & Total edge energy $\leq$ 190\,J/window \\
I-13 & Meet AR latency and task-success targets for real-time service under shared radio and edge load. & AR latency $\leq$ 400\,ms; task success ratio $\geq$ 0.27 \\
I-14 & Maintain streaming throughput while limiting handover instability for mobile users. & Downlink throughput $\geq$ 28\,Mb/s; ping-pong rate $\leq$ 0.05\,s$^{-1}$ \\
I-15 & Preserve the throughput contract while respecting the aggregate AP-energy budget. & Downlink throughput $\geq$ 28\,Mb/s; AP energy $\leq$ 260\,J/window \\
I-16 & Meet AR latency while limiting UE battery consumption during offloaded service. & AR latency $\leq$ 400\,ms; UE energy $\leq$ 0.1\,J/window \\
I-17 & Maintain throughput while preserving reliable packet delivery under radio contention. & Downlink throughput $\geq$ 28\,Mb/s; transmission success $\geq$ 0.6 \\
I-18 & Request ultra-tight AR latency for a mobile interactive session beyond the measured reachable range. & AR latency $\leq$ 50\,ms \\
I-19 & Request aggregate AP energy below the measured reachable level while keeping service available. & AP energy $\leq$ 150\,J/window \\
\bottomrule
\end{tabularx}
\end{table*}

\begin{table*}[!t]
\centering
\caption{Benchmark case compositions and case design.}
\label{tab:casecatalog}
\scriptsize
\renewcommand{\arraystretch}{1.15}
\begin{tabularx}{\textwidth}{@{}
l
>{\raggedright\arraybackslash}p{0.14\textwidth}
>{\raggedright\arraybackslash}p{0.07\textwidth}
X
@{}}
\toprule
\textbf{Case} &
\textbf{Incumbent intents (Assigned xApps)} &
\textbf{Incoming intents} &
\textbf{Case design} \\
\midrule
C-01 &
I-01 (X-06); I-04 (X-10) &
I-08 &
Clean: disjoint KPI contracts and xApps between incumbent and incoming intents; no M1--M5 mechanism intended. \\

C-02 &
I-04 (X-10) &
I-06 &
M1: the incumbent latency policy and the incoming task-success policy can assign distinct xApps acting on $\nu^{\mathrm{cap}}$. \\

C-03 &
I-01 (X-05); I-02 (X-08) &
I-19 &
Negative control: I-19 requests AP energy below the measured reachable level; the expected outcome is rejection. \\

C-04 &
I-01 (X-08 + X-07) &
I-10 &
M2: AP-energy and incumbent throughput xApps act on disjoint attributes; the two KPIs are coupled through radio availability. \\

C-05 &
I-05 (X-09) &
I-12 &
M2: edge-consolidation and incumbent latency xApps act on disjoint attributes; edge energy and latency are coupled through the edge queue. \\

C-06 &
I-04 (X-12); I-06 (X-15); I-08 (X-16) &
I-01, I-05, I-07, I-17, I-18 &
M1: I-05 latency candidates vs.\ $\nu^{\mathrm{cap}}$ of incumbent I-06; I-01/I-07/I-17 candidates vs.\ $w_{\mathrm{5G}}$ of incumbent I-04. M2: radio and edge coupling among throughput, latency, and task KPIs. M4: I-17 may reuse the X-16 instance of incumbent I-08. Negative control: I-18. \\

C-07 &
I-08 (X-16); I-09 (X-17); I-10 (X-02); I-11 (X-03) &
I-13, I-14, I-15, I-17 &
M1: I-13 latency candidates vs.\ $\nu^{\mathrm{cpu}}$ of incumbent I-11; I-14 mobility candidates vs.\ $b_{\mathrm{5G}}$ and $H^{\mathrm{ho}}$ of incumbent I-09. M2: KPI coupling through the shared RAN state. M4: I-17, I-14, and I-15 may reuse the incumbent X-16, X-17, and X-02 instances. M5: I-15 throughput vs.\ the I-10 AP-energy budget through radio sleep availability. \\

C-08 &
I-01 (X-05); I-04 (X-09); I-06 (X-15); I-08 (X-16) &
I-03, I-07, I-14, I-16, I-18 &
M1: I-16 latency candidates vs.\ $\nu^{\mathrm{cap}}$ of incumbent I-06; throughput (I-03, I-14), latency (I-16), and task-success (I-07) candidates can contend on $w_{\mathrm{5G}}$. M3: AP or edge runtime state without direct $W_x$ or $T_x$ overlap. M4: I-03 and I-07 may reuse the incumbent X-05 and X-15 instances. M5: the I-16 UE-energy limit suppresses the edge offloading on which the I-04 latency contract relies. Negative control: I-18. \\

C-09 &
I-01 (X-05); I-06 (X-15); I-08 (X-16); I-09 (X-17); I-10 (X-02) &
I-13, I-14, I-15, I-17, I-18 &
M1: I-13 latency candidates vs.\ $\nu^{\mathrm{cap}}$ of incumbent I-06; I-14 mobility candidates vs.\ $b_{\mathrm{5G}}$ and $H^{\mathrm{ho}}$ of incumbent I-09. M2: KPI coupling through the shared RAN state. M4: I-17, I-14, and I-15 may reuse the incumbent X-16, X-17, and X-02 instances. M5: I-15 throughput vs.\ the I-10 AP-energy budget. Negative control: I-18. \\

C-10 &
I-01 (X-05); I-06 (X-15); I-08 (X-16); I-09 (X-17); I-10 (X-02) &
I-05, I-12, I-14, I-18, I-19 &
M1: I-14 mobility candidates vs.\ $b_{\mathrm{5G}}$ and $H^{\mathrm{ho}}$ of incumbent I-09; I-05 latency candidates vs.\ $\nu^{\mathrm{cap}}$ of incumbent I-06; I-05 and I-14 candidates can contend on $w_{\mathrm{5G}}$. M2: KPI coupling through the shared RAN state. M3: I-12 edge consolidation can violate latency without $W_x$/$T_x$ overlap. M4: I-14 may reuse the incumbent X-17 instance. M5: the I-12 edge-energy budget vs.\ the I-05 latency contract. Negative control: I-18 and I-19. \\
\bottomrule
\end{tabularx}
\vspace{-0.1cm}
\end{table*}

\vspace{-0.3cm}
\subsection{xApp, Intent, and Case Suite}
\label{sec:results:suite}
To evaluate rApp policy orchestration over heterogeneous xApps, we use the 17-xApp catalog in Table~\ref{tab:xappcatalog}. The catalog covers the network attributes in~\eqref{eq:env:control_vector}. It spans energy management, throughput optimization, latency control, edge admission, task success, transmission success, and mobility stability, and its algorithms include random forests (RF), linear upper confidence bound (LinUCB), multilayer perceptrons (MLP), deep Q-networks (DQN), quantile gradient-boosted decision trees (QGBDT), and MLP with action guard (MLP-G). The framework can activate, share, assign, or avoid xApps in the catalog. Fig.~\ref{fig:xapp_controller_validation} verifies that each xApp produces target-direction KPI improvement when evaluated alone against the bare-network baseline.

The experiments evaluate 19 operator intents, summarized in Table~\ref{tab:intentcatalog}. The first group, I-01--I-12, contains single-KPI intents for downlink throughput, AR latency, task success, transmission success, ping-pong rate, AP energy, and edge energy. The second group, I-13--I-17, combines KPIs that are individually feasible but harder to satisfy jointly, such as throughput with mobility stability, throughput with AP energy, latency with UE energy, and throughput with transmission success. The final two intents, I-18 and I-19, are outside the measured reachable KPI ranges. They are designed to test rejection behavior and prevent unconditional xApp deployment from inflating the admission ratio.

The case suite in Table~\ref{tab:casecatalog} evaluates these intents and xApps under incumbent/incoming deployment settings, reflecting network operation in which incoming intents are considered alongside incumbent intents already served by deployed rApp policies.
The cases are ordered from simple to difficult according to two design axes: the number of incumbent and incoming intents and the conflict mechanisms M1--M5 represented in each case. C-01 checks clean admission with disjoint KPI contracts and xApps. C-02--C-05 isolate one dominant deployment condition at a time, including shared network attributes, coupled KPI effects, and a single infeasible incoming request. C-06--C-10 combine multiple incumbents, four or five incoming intents, and mixed conflict mechanisms. C-06, C-08, and C-09 additionally include the infeasible negative-control intent I-18, while C-10 includes both I-18 and I-19.

\begin{figure*}[t]
\centering
\includegraphics[width=\textwidth]{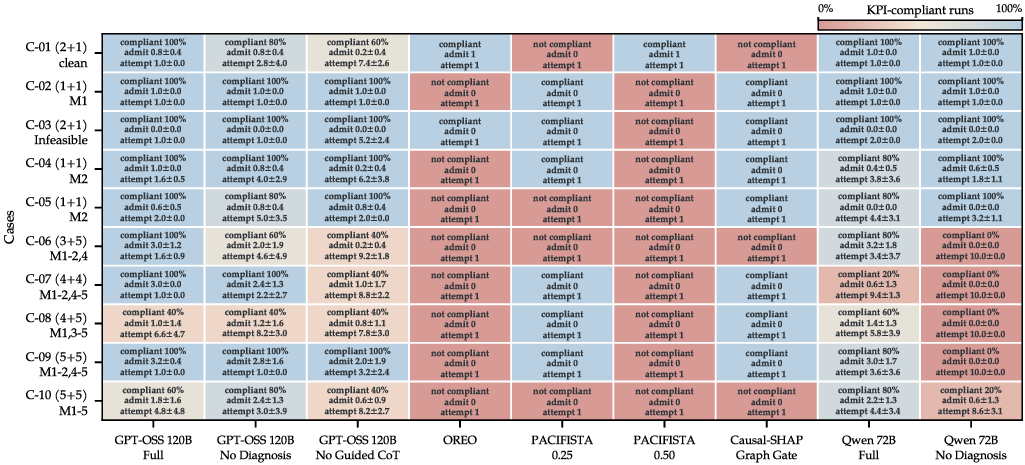}
\caption{Performance comparison across the 10 deployment cases in Table~\ref{tab:casecatalog}. In each row label, $(a+b)$ gives the number of incumbent and incoming intents, and the second line lists the conflict mechanisms or infeasible-request marker represented by the case. Cells for the proposed solution, its ablations, and local LLM variants report the percentage of KPI-compliant runs and the mean $\pm$ standard deviation of admitted incoming intents and deployment attempts across five runs. Baseline cells report the compliance outcome, admitted incoming-intent count, and deployment-attempt count from one run.}
\label{fig:case_solution_outcome_heatmap}
\end{figure*}

\vspace{-0.3cm}
\subsection{Benchmark Configurations}
\label{sec:results:configs}

To distinguish overall performance, component contributions, and the effect of the local LLM backend, the experiments compare closed-loop orchestration with one-shot baselines, internal ablations, and local LLM variants.

\noindent \textit{1) OREO:}
We implement an OREO-based baseline~\cite{mungari2025ran} that preserves priority-weighted service admission and tradeoffs involving xApp sharing and resource cost under the same KPI validation procedure. Because our setting is defined by KPI contracts and xApp catalog metadata rather than O-Cloud CPU/RAM/disk placement or explicit service-function graphs, we adapt OREO's service-configuration selection by mapping each target KPI to candidate xApps from the xApp catalog. The baseline filters configurations that violate KPI contracts and greedily selects the feasible configuration with the highest adapted score. This score combines intent priority, the cost of newly instantiated xApps, reuse of already active xApps, and KPI margin.

\noindent \textit{2) PACIFISTA:}
A PACIFISTA-based baseline~\cite{delprever2025pacifista} is constructed from profiles collected in the O-RAN environment. Each xApp is executed alone to collect observations for empirical cumulative distribution functions (ECDFs) over network attributes and KPI measurements. These profiles, together with xApp catalog metadata and latent-resource links, form the augmented application--attribute--KPM conflict graph. The KPM severity term $\sigma_K$ is computed from ECDF integral-area distances over the measurements used to evaluate the corresponding KPI contracts. For each incoming intent's fixed candidate xApp set, the admission score is the maximum $\sigma_K$ between any candidate xApp and the currently deployed xApp set. Incoming intents are processed greedily by priority and admitted only when this score is at most $\delta^{\mathrm{TOL}}$. Following the tolerance examples in the reference, we report $\delta^{\mathrm{TOL}}=0.25$ and $\delta^{\mathrm{TOL}}=0.50$.

\noindent \textit{3) Causal-SHAP Graph Gate:}
The Causal-SHAP graph gate follows the XGBoost-SHAP/DoWhy causal-DAG path of Sharma \textit{et al.}~\cite{sharma2025towards} on the same xApp profile data. SHAP importance over per-KPI XGBoost surrogates retains attribute--KPI edges, and DoWhy backdoor regression estimates average treatment effects and produces refutation outputs for those retained edges. We then convert the learned attribute--KPI graph into a conservative xApp-set admission score. An incoming intent is greedily admitted only when the risk score of its candidate xApp set against the currently deployed xApps remains below the tolerance.

\noindent \textit{4) Agentic ablations and local LLM variants:}
The agentic configurations use locally served LLMs so that operator intents, incumbent deployment state, and KPI telemetry remain within the local orchestration environment. The default backend is GPT-OSS 120B. GPT-OSS Full uses the complete Reasoning--Diagnosis loop described in Section~\ref{sec:agentai}. GPT-OSS No Diagnosis keeps the same Reasoning Agent, supervision module, and retry budget, but removes Diagnosis feedback from retry attempts. GPT-OSS No Guided CoT keeps both agents and the same input/output contracts, but replaces the guided reasoning prompts with objective-only role prompts. Qwen 72B Full and Qwen 72B No Diagnosis keep the same orchestration solution and validation procedure and change only the local LLM backend. All agentic configurations allow at most 10 deployment attempts, consisting of one initial attempt and $K_r=9$ retries. A budget-exhausted cycle reaches this limit without finding a KPI-compliant deployment. For consistent reporting, cycles ending in a timeout, interruption, or runtime error are assigned 10 deployment attempts and zero admitted incoming intents.

\vspace{-0.3cm}
\subsection{Results}
\label{sec:results:main}
To evaluate overall orchestration performance while accounting for LLM output variability, we ran each case five times for the full Agentic AI solution, its ablations, and the local LLM variants. Fig.~\ref{fig:case_solution_outcome_heatmap} reports the per-case KPI-compliant run percentage and the mean and standard deviation of admitted incoming intents and deployment attempts for these Agentic AI experiments, together with the single-run baseline outcomes. A case outcome is KPI-compliant only when all incumbent intents and all admitted incoming intents satisfy their KPI contracts after deployment. As shown in the figure, GPT-OSS 120B Full is compliant across all five runs in 8 of the 10 cases and has a nonzero mean admitted-intent count in every case except the infeasible C-03. In the larger cases C-06, C-07, and C-09, it remains compliant in all five runs while admitting 3.0, 3.0, and 3.2 incoming intents on average. By comparison, Causal-SHAP and PACIFISTA with $\delta^{\mathrm{TOL}}=0.25$ reach 7 and 6 compliant cases but admit no incoming intents. OREO and PACIFISTA with $\delta^{\mathrm{TOL}}=0.50$ each admit only 1 incoming intent and reach 2 and 1 compliant cases, respectively. This difference could be attributed to the one-shot decision structure of the baselines. Their decisions rely on pre-deployment evidence, so compliance is achieved mainly by rejecting incoming intents instead of revising xApp assignments after KPI supervision.

The No Guided CoT ablation examines the contribution of the guided reasoning order while retaining both agents and the same input/output contracts. Table~\ref{tab:agentic_summary} shows that the ablation averages 7.2 compliant cases, 6.8 admitted incoming intents, and 5.90 attempts per case, compared with 9.0, 15.4, and 2.16 for GPT-OSS 120B Full. Comparing the corresponding columns in Fig.~\ref{fig:case_solution_outcome_heatmap}, the compliant-run percentage falls from 100\% to 40\% in both C-06 and C-07. A similar gap appears in the attempt counts for cases where both configurations are fully compliant. No Guided CoT requires 6.2 versus 1.6 attempts in C-04 and 3.2 versus 1.0 in C-09. These results show that the guided CoT stage order helps the agents convert conflict evidence and KPI feedback into effective screening and assignment revision across attempts. 
Table~\ref{tab:agentic_summary} also summarizes the performance of the two local LLMs under the same orchestration logic. Of the two, Qwen 72B Full achieves 7.8 compliant cases, 12.8 admitted incoming intents, and 3.88 attempts per case, whereas GPT-OSS 120B Full reaches 9.0, 15.4, and 2.16, respectively. Fig.~\ref{fig:case_solution_outcome_heatmap} provides a case-level example of this gap. In C-07, GPT-OSS is compliant in all five runs, while Qwen is compliant in one. Together, the aggregate and case-level results highlight the importance of LLM reasoning capability in conflict-aware orchestration.

Finally, the Full and No Diagnosis comparison isolates the contribution of post-deployment failure attribution under the same Reasoning Agent, supervision module, and deployment-attempt budget. Table~\ref{tab:agentic_summary} shows a smaller difference with GPT-OSS 120B. GPT-OSS 120B Full reaches 9.0 compliant cases and 15.4 admitted incoming intents, compared with 8.4 and 14.2 for No Diagnosis, while reducing attempts from 3.28 to 2.16. The difference becomes larger with Qwen 72B. Diagnosis increases the averages from 5.2 to 7.8 compliant cases and from 3.2 to 12.8 admitted incoming intents, while reducing attempts from 5.76 to 3.88. This widening gap shows that Diagnosis becomes more important as the LLM's reasoning capability decreases. In this setting, the Reasoning Agent relies more strongly on structured failure attribution to identify an effective revision for the next assignment.

\begin{table}[t]
\centering
\caption{Aggregate performance over the 10 deployment cases and five repeated runs. Values are reported as mean $\pm$ standard deviation across runs.}
\label{tab:agentic_summary}
\scriptsize
\setlength{\tabcolsep}{2.2pt}
\renewcommand{\arraystretch}{1.08}
{\renewcommand{\tabularxcolumn}[1]{m{#1}}
\begin{tabularx}{\columnwidth}{@{}
>{\centering\arraybackslash}m{0.18\columnwidth}
>{\raggedright\arraybackslash}m{0.22\columnwidth}
>{\centering\arraybackslash}X
>{\centering\arraybackslash}X
>{\centering\arraybackslash}X
@{}}
\toprule
\makecell[c]{Model} &
\makecell[c]{Variant} &
\makecell[c]{KPI-compliant\\cases} &
\makecell[c]{Admitted\\intents} &
\makecell[c]{Attempts\\per case} \\
\midrule
\multirow{3}{*}{\makecell[c]{GPT-OSS\\120B}}
& Full
& 9.0$\pm$1.0
& 15.4$\pm$3.8
& 2.16$\pm$0.91 \\
& No Diagnosis
& 8.4$\pm$1.8
& 14.2$\pm$3.8
& 3.28$\pm$1.06 \\
& No Guided CoT
& 7.2$\pm$0.8
& 6.8$\pm$2.2
& 5.90$\pm$0.68 \\
\midrule
\multirow{2}{*}{\makecell[c]{Qwen\\72B}}
& Full
& 7.8$\pm$0.8
& 12.8$\pm$3.6
& 3.88$\pm$0.44 \\
& No Diagnosis
& 5.2$\pm$0.4
& 3.2$\pm$1.6
& 5.76$\pm$0.37 \\
\bottomrule
\end{tabularx}}
\vspace{0cm}
\end{table}

\vspace{-0.3cm}
\section{Proof of Concept}
\label{sec:poc}
Complementing the evaluation in Section~\ref{sec:results}, we deploy the proposed framework as a proof of concept (PoC) on a live over-the-air O-RAN testbed. The testbed connects the full orchestration loop to real radio hardware for multi-intent rApp policy orchestration on a single live cell serving commercial user equipment (UEs).

\vspace{-0.3cm}
\subsection{Testbed Setup}
\label{sec:poc_setup}
\begin{figure}[!t]
    \centering
    \includegraphics[width=1\linewidth]{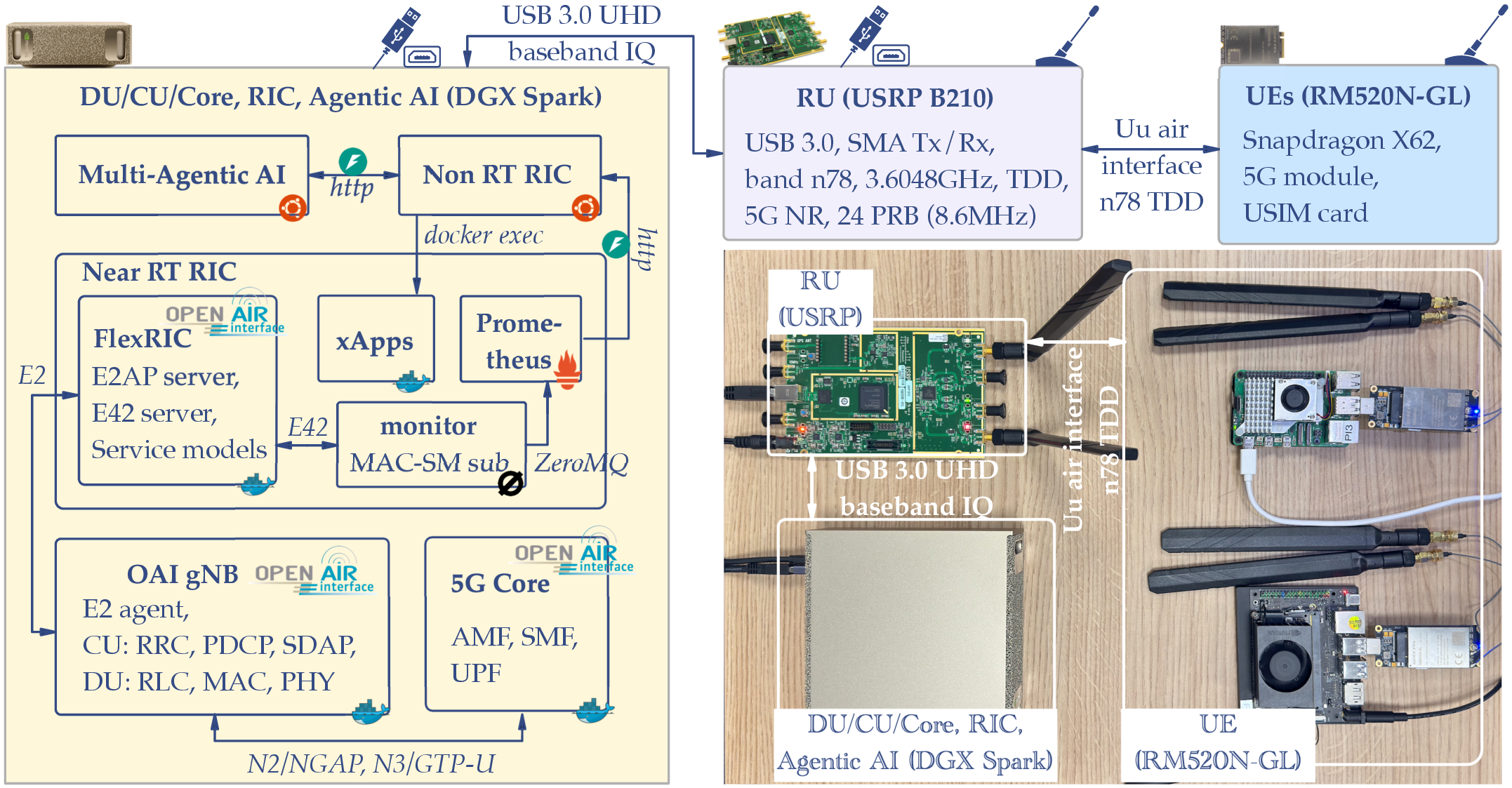}
    \caption{Over-the-air O-RAN testbed with a softwarized DU, CU, and 5G core on a single host, a USRP B210 RF front-end, and two commercial 5G modules as UEs.}
    \label{fig:poc_testbed}
    \vspace{-0.2cm}
\end{figure}

The PoC setup is shown in Fig.~\ref{fig:poc_testbed}. The OpenAirInterface (OAI) distributed unit (DU), centralized unit (CU), and 5G core run in Docker containers on a DGX Spark, with an Ettus USRP B210 as the radio unit (RU). The band n78 cell operates at 3.6048\,GHz with time-division duplexing (TDD), 30\,kHz subcarrier spacing, and 24 physical resource blocks (PRBs). Two Quectel RM520N-GL modules attach over the air as UEs. FlexRIC connects to the E2 agent over E2AP/SCTP using the KPM, MAC, and SLICE service models (SMs).

For monitoring, a MAC-SM monitor xApp streams per-UE throughput, modulation and coding scheme (MCS), block error rate (BLER), PRB usage, and signal-to-noise ratio through ZeroMQ into Prometheus. A DU-side scheduler extension exposes three per-UE network attributes: PRB quota, MCS cap, and proportional-fair scheduling weight. Python xApps built on the FlexRIC SDK subscribe to MAC-SM telemetry and apply these attributes through SLICE-SM control messages over the standard xApp $\rightarrow$ Near-RT RIC $\rightarrow$ E2AP $\rightarrow$ DU path. 
The Agentic AI layer and the Non-RT RIC function are implemented as separate components. The former reuses the Reasoning and Diagnosis Agents of Section~\ref{sec:agentai}, including their guided CoT prompts and conflict taxonomy, with GPT-OSS 120B running on the DGX Spark. The Non-RT RIC function maintains the xApp catalog and incumbent deployment state and launches the xApps selected by each candidate policy.

\vspace{-0.3cm}
\subsection{xApp Catalog, Intents, and Cases}
\label{sec:poc_catalog}

\begin{table}[!t]
\centering
\caption{Operator intents used in the PoC. DL denotes downlink.}
\label{tab:poc_intents}
\scriptsize
\setlength{\tabcolsep}{3pt}
\renewcommand{\arraystretch}{1.08}
\begin{tabularx}{\columnwidth}{@{}
>{\raggedright\arraybackslash}p{0.07\columnwidth}
>{\raggedright\arraybackslash}X
>{\raggedright\arraybackslash}p{0.4\columnwidth}
@{}}
\toprule
\textbf{Intent} &
\textbf{Operator request} &
\textbf{KPI contract} \\
\midrule
LI-01 &
Limit a target UE's downlink to 2\,Mb/s &
$|\mathrm{DL} - 2.0| \leq 0.25$\,Mb/s \\

LI-02 &
Guarantee a target UE a 4.5\,Mb/s downlink &
DL $\geq$ 4.5\,Mb/s \\

LI-03 &
Cap a target UE's downlink MCS at 4 &
Window-average DL MCS $\leq$ 4 \\

LI-04 &
Split the band into two 12-PRB partitions &
$\approx$12 PRBs per UE \\

LI-05 &
Guarantee an 8\,Mb/s downlink &
DL $\geq$ 8\,Mb/s (infeasible) \\

LI-06 &
Guarantee a target UE a 3\,Mb/s downlink &
DL $\geq$ 3\,Mb/s \\
\bottomrule
\end{tabularx}
\end{table}

Five xApps were developed for the PoC: (i) the \emph{PRB-Rate-Limiter} adjusts a UE's PRB quota every 3\,s toward a throughput target; (ii) the \emph{MCS-Governor} monitors DL BLER and adapts the maximum DL MCS to trade spectral efficiency for link robustness; (iii) the \emph{Admission-Control} xApp suspends or resumes downlink scheduling; (iv) the \emph{PRB-Splitter} partitions the 24-PRB band between two UEs; and (v) the \emph{Attribute-Controller} applies a one-shot setting to the PRB quota, MCS cap, or scheduling weight.
The six live-testbed intents in Table~\ref{tab:poc_intents} are evaluated across five cases that progress from clean admission and pre-deployment conflict screening to runtime coexistence and mixed admission decisions. LC-01 contains no incumbent intent and submits LI-01 to evaluate the complete path from intent submission to xApp deployment and KPI compliance. LC-02 submits LI-01 and LI-06 concurrently for the same UE. Their throughput contracts are individually feasible but mutually incompatible, isolating M5 and testing partial admission within a joint decision. LC-03 places the MCS-cap intent LI-03 against an incumbent rate-limiting policy serving LI-01. Their catalog profiles appear structurally disjoint, so the case probes coexistence behavior that cannot be determined from pre-deployment structural information alone. LC-04 submits the infeasible LI-05 as a negative control to test rejection before deployment. LC-05 combines four incoming intents with incumbent LI-01 to evaluate a mixed decision involving feasible LI-03, M1 contention from LI-04, M5 incompatibility between LI-02 and incumbent LI-01, and the infeasible LI-05. Each incumbent policy is established through the orchestration loop and verified as KPI-compliant before the incoming requests are evaluated.

\begin{table}[!t]
\centering
\caption{Live-testbed orchestration outcomes and their mapping to the conflict taxonomy. Neg.\ denotes the negative control.}
\label{tab:poc_outcomes}
\scriptsize
\setlength{\tabcolsep}{3pt}
\renewcommand{\arraystretch}{1.08}
\begin{tabularx}{\columnwidth}{@{}
>{\raggedright\arraybackslash}p{0.085\columnwidth}
>{\centering\arraybackslash}p{0.1\columnwidth}
>{\raggedright\arraybackslash}p{0.32\columnwidth}
>{\centering\arraybackslash}p{0.03\columnwidth}
>{\raggedright\arraybackslash}X
@{}}
\toprule
\textbf{Case} &
\textbf{Mech.} &
\textbf{Agent Decision} &
\textbf{Att.} &
\textbf{Over-the-Air Evidence} \\
\midrule
LC-01 &
--- &
admit LI-01 with PRB-Rate-Limiter &
1 &
DL $4.6 \rightarrow 2.1$\,Mb/s \\

LC-02 &
M5 &
admit LI-01; reject LI-06 &
1 &
$\leq$2 and $\geq$3 both reachable but incompatible; LI-06 is rejected \\

LC-03 &
M3 &
reject LI-03 after Diagnosis-guided retry &
3 &
DL MCS remains 9; incumbent DL $\approx2.1$\,Mb/s \\

LC-04 &
Neg. &
reject LI-05 as unreachable &
1 &
8\,Mb/s exceeds reachable $\approx5.1$\,Mb/s \\

LC-05 &
M1, M5, Neg. &
admit LI-03; reject LI-04, LI-02, and LI-05 &
1 &
LI-03 compliant; incumbent remains near 2\,Mb/s; LI-02 is individually reachable \\
\bottomrule
\end{tabularx}
\end{table}

\begin{figure}[!t]
    \centering
    \includegraphics[width=0.98\linewidth]{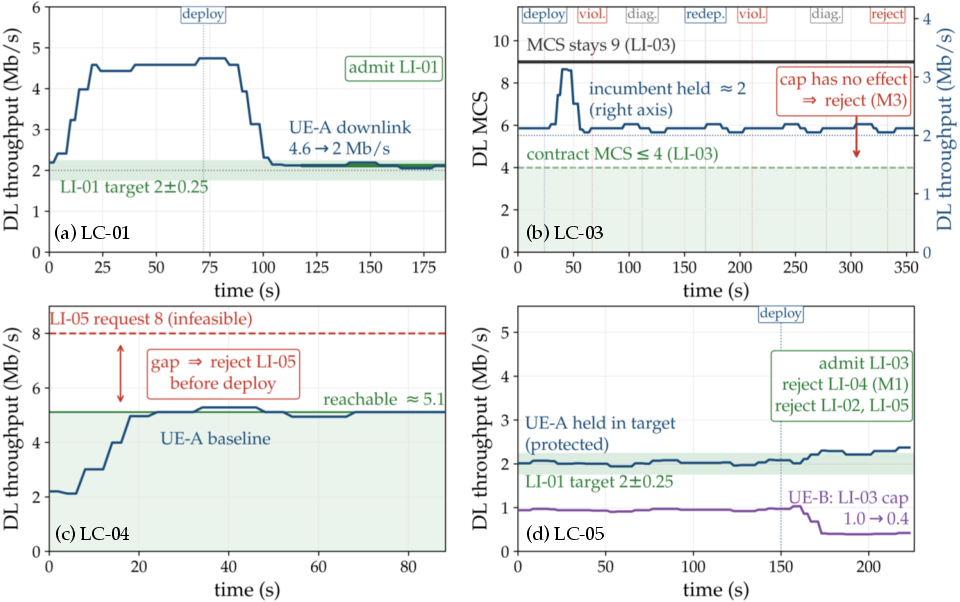}
    \caption{Representative over-the-air PoC timelines. (a) LC-01: LI-01 reaches its throughput contract. (b) LC-03: an M3 runtime coupling prevents the MCS cap from taking effect, while incumbent throughput is preserved. (c) LC-04: the requested 8\,Mb/s exceeds the measured reachable value. (d) LC-05: LI-03 is admitted while M1-conflicting, M5-incompatible, and unreachable requests are rejected.}
    \label{fig:poc_cases}
\end{figure}

\vspace{-0.3cm}
\subsection{Validation Results}
\label{sec:poc_results}
Table~\ref{tab:poc_outcomes} summarizes all five outcomes, while Fig.~\ref{fig:poc_cases} shows representative over-the-air timelines. In LC-01, the Reasoning Agent admits LI-01 and assigns the PRB-Rate-Limiter. Its closed-loop PRB control reduces downlink throughput from 4.6\,Mb/s into the $2.0\pm0.25$\,Mb/s contract band, with a post-convergence window average of 2.12\,Mb/s (Fig.~\ref{fig:poc_cases}(a)).
LC-02 presents an M5 conflict between LI-01 and LI-06, whose throughput contracts cannot be satisfied simultaneously. The solution admits LI-01 and rejects LI-06, retaining one feasible intent from the joint request.
LC-03 triggers post-deployment diagnosis (Fig.~\ref{fig:poc_cases}(b)). The first assignment combines the incumbent PRB-Rate-Limiter with the MCS-Governor, whose catalog profiles declare disjoint controlled attributes and affected KPI sets. After deployment, the window-average MCS remains 9, violating the MCS cap of 4. Diagnosis attributes the violation to an M3 runtime coupling in which both xApps update the same E2 slice-control object and the incumbent's periodic refresh erases the MCS cap. An isolated validation confirms that the cap itself is effective, reducing MCS from 9 to 4 while sustaining 2.5\,Mb/s. Reassignment to the Attribute-Controller also leaves the MCS at 9. Following these unsuccessful assignments, the Reasoning Agent rejects LI-03 on the third attempt under the persistent-failure rule without another deployment, while incumbent throughput remains near 2.1\,Mb/s.
LC-04 rejects LI-05 because its requested 8\,Mb/s exceeds the measured reachable value of approximately 5.1\,Mb/s (Fig.~\ref{fig:poc_cases}(c)). In LC-05, the Reasoning Agent admits LI-03 on the second UE, rejects LI-04 because of M1 overlap with the incumbent PRB writer, rejects LI-02 because its individually reachable 4.5\,Mb/s contract is incompatible with incumbent LI-01 under M5, and rejects the infeasible LI-05. KPI supervision confirms that LI-03 is compliant, with the second UE's downlink throughput decreasing from 1.0 to 0.4\,Mb/s while the incumbent remains near 2\,Mb/s (Fig.~\ref{fig:poc_cases}(d)).

\vspace{-0.2cm}
\section{Conclusion}
\label{sec:conclusion}
This paper presented a closed-loop Agentic AI framework for intent-driven and conflict-aware rApp policy orchestration in O-RAN. The framework jointly decides incoming-intent admission and xApp assignment, with each admitted intent realized by an rApp policy composed of assigned xApps. The framework combines a guided CoT Reasoning Agent for pre-deployment screening and assignment, a guided CoT Diagnosis Agent for post-deployment failure attribution, and structured non-LLM modules for policy deployment and KPI supervision. 
Across five runs per case, the proposed solution based on GPT-OSS 120B achieves 28.6\% more KPI-compliant cases and admits over 15$\times$ as many incoming intents as state-of-the-art baselines. The ablations show that guided CoT improves compliant admission and attempt efficiency, while the model comparison highlights the importance of LLM reasoning capability in mixed-conflict cases. Diagnosis partly compensates for reduced reasoning capability, increasing Qwen 72B's compliant cases by 50.0\% and reducing mean deployment attempts by 32.6\%. A live over-the-air OAI/FlexRIC proof of concept further demonstrates the practical feasibility of the framework by preserving incumbent-intent compliance while orchestrating rApp policies in real networks.

\bibliographystyle{IEEEtran}
\bibliography{references}

\end{document}